\documentclass{ifacconf}

\usepackage{graphicx}      
\usepackage{natbib}       
\usepackage{amsmath} 
\usepackage{graphics}
\usepackage[dvipsnames]{xcolor}
\usepackage{svg}
\usepackage{textcomp}
\begin{document}
\begin{frontmatter}
\title{Robust Iterative Learning for Collaborative Road Profile Estimation and Active Suspension Control in Connected Vehicles\thanksref{footnoteinfo}} 

\thanks[footnoteinfo]{This work is supported by the U.S. National Science Foundation (Grants: No. 2030375,  2030411, and 2422579)}

\author[First]{Harsh Modi} 
\author[Second]{Mohammad R Hajidavalloo} 
\author[Third]{Zhaojian Li}
\author[Fourth]{Minghui Zheng}

\address[First]{Texas A\&M University, College Station 77840, TX, USA(e-mail: harsh.modi@tamu.edu).}
\address[Second]{Michigan State University, East Lansing, MI 48824, USA (e-mail: hajidava@msu.edu)}
\address[Third]{Michigan State University, East Lansing, MI 48824, USA (e-mail: lizhaoj1@egr.msu.edu)}
\address[Fourth]{{Texas A\&M University, College Station 77840, TX  (e-mail: mhzheng@tamu.edu).}}

\begin{abstract} 
This paper presents the development of a new collaborative road profile estimation and active suspension control framework in connected vehicles, where participating vehicles iteratively refine the road profile estimation and enhance suspension control performance through an iterative learning scheme. Specifically, we develop a robust iterative learning approach to tackle the heterogeneity and model uncertainties in participating vehicles, which are important for practical implementations. In addition, the framework can be adopted as an add-on system to augment existing suspension control schemes. Comprehensive numerical studies are performed to evaluate and validate the proposed framework.
\end{abstract}


\end{frontmatter}

\section{Introduction}

There is a growing interest in utilizing road profile information to enhance suspension control with improved comfort and safety [\cite{active_suspension_advances,Mohammad_TITS}]. The estimated road profile information can also be used for optimized budget allocation for pavement maintenance [\cite{pavement_assessment_for_decision_making}]. Traditionally, the road profile estimation process involves specialized sensors such as contact-based  [\cite{contact_method1}, \cite{contact_method2}] and laser-based sensors [\cite{laser_method1}, \cite{laser_method2}], which are costly to acquire and maintain and can only provide limited coverage [\cite{comfort}].

On the other hand, modern vehicles are equipped with a multitude of sensors [\cite{cars_have_sensors}], which can be readily integrated with advanced machine learning and communication telematics for efficient road data crowdsourcing [\cite{autonomous_vehicles_survey,vehicular_cloud_computing}]. The crowdsourced data has been proven effective in accurately estimating road and traffic information such as real-time traffic data (e.g. Google Maps, Waze). With advancements in onboard vehicular sensing and communication, the process of estimating the road profile can also take advantage of the crowdsourced data. Specifically, with multiple participating vehicles estimating and sharing the information on the same road segment, vehicle-specific biases and errors can be mitigated and widespread road profile estimations can be done in a cost-effective manner.

The onboard sensor-based road profile estimation has been done in some studies. \cite{hinf_onboard} utilized H-infinity based observer and onboard sensors to estimate the road profile in the passive suspension vehicles. \cite{semi_active_Q_parameterization} utilizes the Q-parametrization approach to estimate the road profile for semi-active suspension systems. \cite{active_sensor_front} utilizes the onboard sensor capable of measuring the road height profile in front of the vehicle to control the active suspension force. \cite{active_MPC} utilized model predictive control to predict the road profile using the lead vehicle preview. Some other studies used methodologies such as observer and dynamic response of the vehicle [\cite{previous1}, \cite{previous2},\cite{previous3}]. However, all of these research methods were focused on using only one vehicle, which is prone to inaccuracies due to vehicle-specific characteristics. 

\begin{figure}[!htbp]
         \centering {\includegraphics[width=0.38\textwidth]{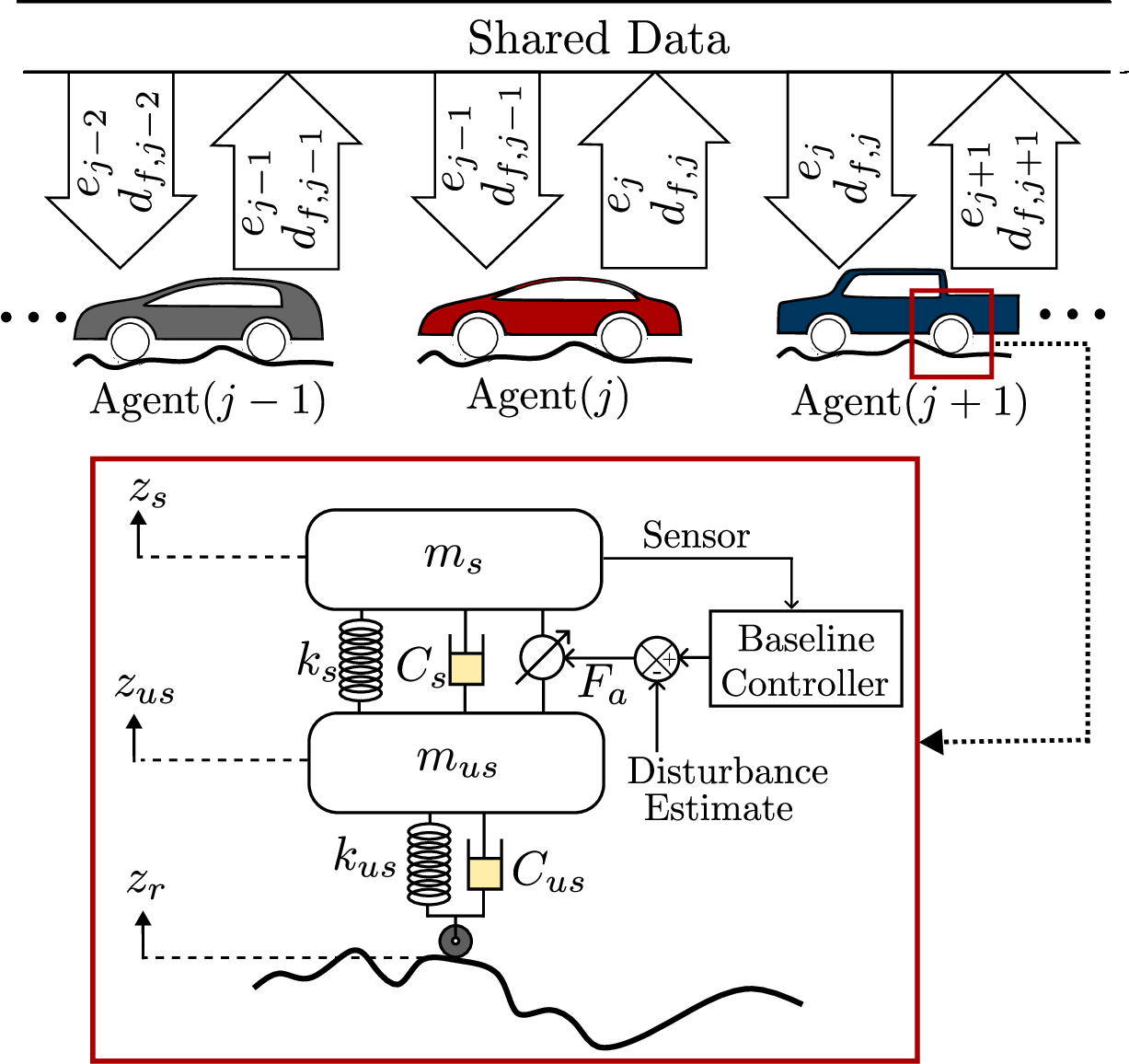}}
         \caption{\textcolor{black}{Overall learning schematic along with a quarter car model used for one of the agents}}
      \label{fig:framework}
\end{figure}

There exist some studies utilizing disturbance observer (DOB) and iterative learning control to improve disturbance estimate and trajectory tracking for drones [\cite{chen2020knowledge}, \cite{zheng2020generalized}, \cite{previous4}], which has not been investigated in active suspension. Existing studies on using multiple vehicles and iterative learning to improving road profile estimation [\cite{chen2022cascaded}, \cite{gao2020privacy}] mainly focus on passive suspension systems without closed-loop feedback. Learning-based approach with an active suspension system can estimate the road profiles more accurately and robustly as closed-loop feedback inherently ensures the correction of estimation errors. In this study, we develop a new method that leverage DOB and iterative learning control for estimating the road profile using connected vehicles. To the best of our knowledge, this is the first attempt at utilizing closed-loop active suspension with learning among multiple vehicles to estimate the road profiles with explicitly considering vehicle modeling uncertainties.

In this research, the learning is among the vehicles with different dynamics. The difference in dynamics can arise either from different type of vehicles used or from the differences in payload, tire pressure, suspension coefficients, etc. within the same vehicle type. The learning framework is designed to account for these differences. The basic outline of the learning framework is as follows: 
\vspace{-5pt}
\begin{enumerate}
    \item Each vehicle equipped with active suspension estimates the road profile on its own using DOB.\\
    \vspace{-10pt}
    \item The vehicle also receives anonymous data from the previously passed vehicle. This data includes information about its nominal dynamics model and its errors. \\
    \vspace{-10pt}
    \item Based on this data, the learning filters generate a learning signal, which is added to the road profile estimate from DOB, improving the estimation accuracy.\\
\end{enumerate}
\vspace{-10pt}
The remaining paper is organized as follows: In section \ref{section:learning_framework}, we will establish the theoretical learning framework. In section \ref{section:simulation}, we will present the numerical studies and we will conclude the article in section \ref{section:conclusion}.

\section{Learning Framework}
\label{section:learning_framework}
\vspace{-7pt}
The learning framework is implemented in a cascaded format, i.e. an agent $\#(j)$ learns from the information of agent $\#(j-1)$ and agent $\#(j)$ passes on the data to be used by agent $\#(j+1)$. Fig. \ref{fig:framework} shows the overall learning information flow among different agents (vehicles). They share information such as their own learning signal, the sprung mass displacement, and their dynamics to a shared database. We use a quarter-car model for developing the learning framework. Fig. \ref{fig:framework} also shows a detailed version of this quarter car model for one of the agents.

In Fig. \ref{fig:framework}, $m_s$ is a sprung mass, $m_{us}$ is an unsprung mass, $k_s$ and $k_{us}$ are suspension stiffness and tire stiffness coefficients respectively, and $C_s$ and $C_{us}$ are spring and tire damping coefficients respectively. The Active suspension force is described as $F_a$. $z_s$, $z_{us}$, and $z_r$ are sprung mass, unsprung mass, and road profile displacements respectively. It is important to note that we treat the road profile $z_r$ as a function of time instead of a function of space. With the speed of the vehicle readily available in the onboard sensing, the road profile estimation can be converted into a spatial domain. In this study, we will consider the road profile as a disturbance. Hence, we subtract the equivalent disturbance estimate from the control signal generated by the baseline controller. This research aims to accurately estimate this disturbance. To establish a disturbance estimation process, let us first describe the governing equations of this quarter-car suspension system:
\begin{small}
\begin{equation}
\label{equation:model_ODE}
\begin{array}{rl}
    m_s \Ddot{z}_s=&F_a + C_s ( \dot{z}_{us}-\dot{z}_s)+k_s(z_{us}-z_s)\\
    m_{us}\Ddot{z}_{us}=&-F_a-C_s(\dot{z}_{us}-\dot{z}_s)-k_s(z_{us}-z_s)\\
    &+C_{us}(\dot{z}_r-\dot{z}_{us})+k_{us}(z_r-z_{us})
\end{array}
\end{equation}
\end{small}
The systems of Eq. (\ref{equation:model_ODE}) have two independent inputs: 1. active suspension force ($F_a$), and 2. the road profile ($z_r$). As we want to minimize the sprung mass displacement ($z_s$), we consider $z_s$ as the output of the system. In this study, we assume can measure $z_s$ using the onboard sensors [\cite{zus_displacement_estimation}]. Using Eq. (\ref{equation:model_ODE}) and assuming the suspension operates in a linear regime, we can derive the transfer functions relating the output ($z_s$) to each of the inputs ($F_a$ and $z_r$) respectively as:
\begin{small}
\begin{equation}
    \label{equation:tranfer_functions1}
    \begin{array}{rc}
    P(1)=z_s/F_a
    =&\frac{m_{us}\cdot s^2+c_{us}\cdot s+k_{us}}{\begin{matrix}
    m_s m_{us}s^4+(c_s m_s+c_s m_{us}+c_{us}m_s)s^3\\
    +(k_s m_s+k_s m_{us}+k_{us}m_s+c_s c_{us})s^2\\
    +(c_s k_{us}+c_{us} k_s)s+k_s k_{us}\end{matrix}}\\
    \end{array}
\end{equation}
\end{small}
\begin{small}
\begin{equation}
    \label{equation:tranfer_functions2}
    \begin{array}{rc}
    P(2)=z_s/\dot{z}_r=&\frac{(k_s+c_s s)(k_{us}+c_{us} s)}{\begin{matrix}
    m_s m_{us}s^5+(c_s m_s+c_s m_{us}+c_{us}m_s)s^4 \\
    +(k_s m_s+k_s m_{us}+k_{us}m_s+c_s c_{us})s^3 \\
    +(c_s k_{us}+c_{us} k_s)s^2+k_s k_{us} s
    \end{matrix}}
    \end{array}
\end{equation}
\end{small}

\begin{figure}[!htbp]
         \centering {\includegraphics[width=0.4\textwidth]{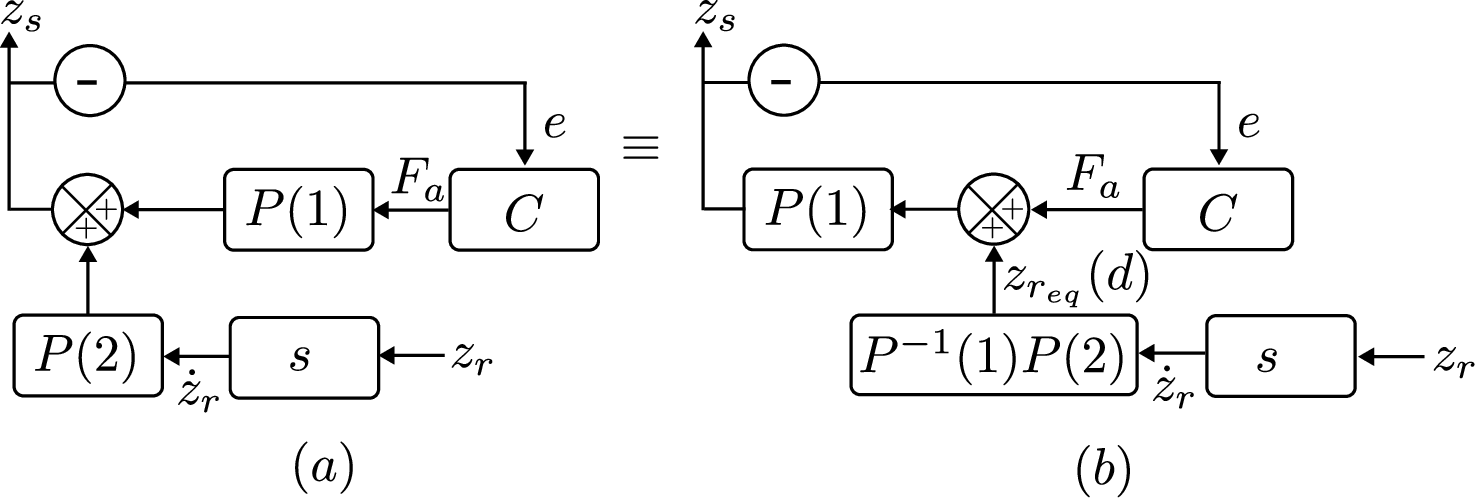}}
         \caption{\textcolor{black}{(a) Original block diagram of the system (b) Equivalent block diagram after block diagram manipulation}}
      \label{fig:system_transfer_function1}
\end{figure}

The transfer function system derived in Eq. (\ref{equation:tranfer_functions1}) and Eq. (\ref{equation:tranfer_functions2}) is arranged in block diagram form in Fig. \ref{fig:system_transfer_function1} (a). In this block diagram, $C$ represents a baseline controller, $e$ represents an error in trajectory tracking (i.e. $-z_s$), and $s$ is a derivative transfer function. The learning framework is developed such that the baseline controller $C$ remains untouched by the framework. Hence, the learning framework can be applied to any active suspension system regardless of what controller is used in the system.

As we consider using a DOB for a primary estimation of the road profile, we need to treat the road profile as a disturbance. Hence, we move the signal corresponding to $z_r$ upstream of P(1) using the block diagram manipulation as shown in Fig. \ref{fig:system_transfer_function1} (b). Considering this, the equivalent disturbance based on the road profile can be given as: 
\begin{small}
\begin{equation}
    \label{equation:zr_eq}
    z_{r_{eq}}=d=\delta\{z_r\} \text{, where   }
    \delta=s P^{-1}(1)P(2)
\end{equation}
\end{small}
As $z_{r_{eq}}$ is being added between the baseline controller and the plant, it can be considered a disturbance to the system. For simplicity, we will represent  $z_{r_{eq}}$ as $d$ in further equations and figures. The notation \{\} means that the signal inside the notation is sent to a system which can be represented by the outside transfer function.

\begin{figure}[htbp]
         \centering {\includegraphics[width=0.3\textwidth]{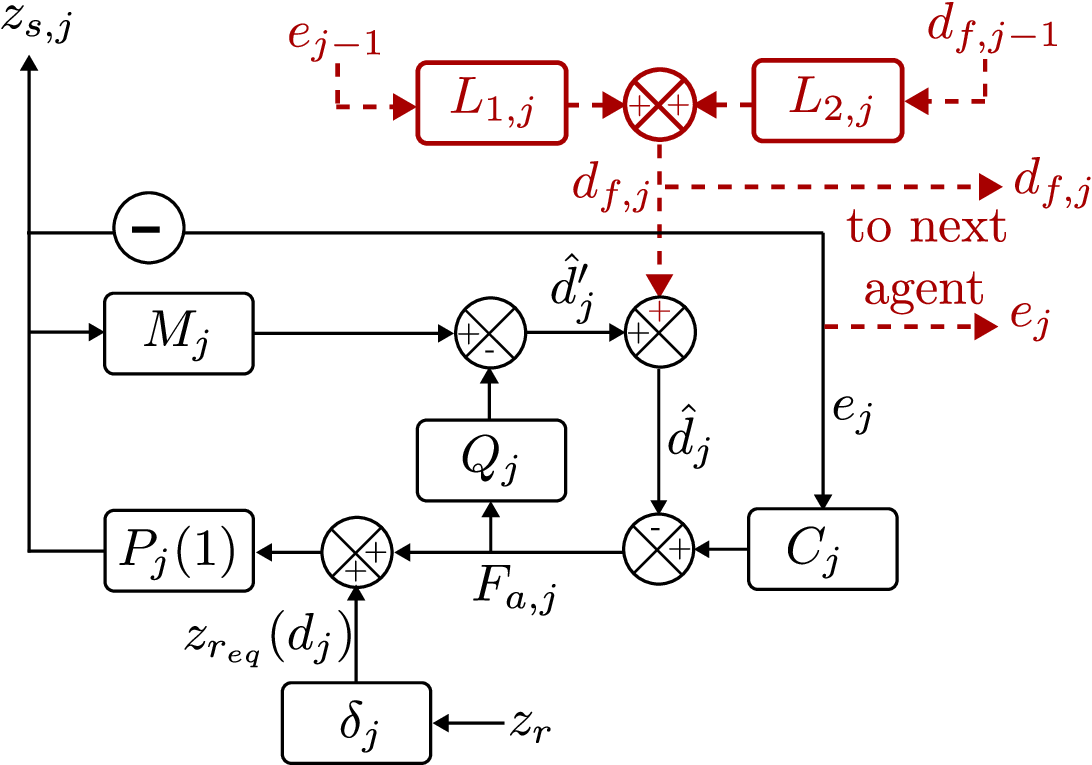}}
         \caption{\textcolor{black}{Block diagram of the system along with learning framework for Agent$\#(j)$}}
      \label{fig:system_transfer_function2}
\end{figure}

Fig. \ref{fig:system_transfer_function2} shows a block diagram of the system with the DOB and learning framework added to the system. As the learning framework establishes the learning relationship between different agents (vehicles), each agent's dynamics can be different from any other agent. In order to denote this difference, we use the index $j$ as the subscript of each term. In Fig. \ref{fig:system_transfer_function2}, $M_j\approx Q_jP^{-1}_j$ is an inverse approximation of the plant $P_j(1)$, $Q_j$ is a low pass filter, $F_{a,j}$ is an active suspension actuator force. $\hat{d}'_j$ is a disturbance estimate from the DOB. We complement this estimate with the learning signal $d_{f,j}$ in order to generate the final disturbance estimate $\hat{d}_j$. This learning signal is generated using the trajectory tracking error of the previous agent ($e_{j-1}$), the learning signal of the previous agent ($d_{f,j-1}$), and respective to-be-designed learning filters $L_{1,j}$, and $L_{2,j}$. Similarly, the data from the current agent is passed to the next agent.

With this agent description established, we will derive the learning filters $L_{1,j}$, and $L_{2,j}$ in the following subsections such that disturbance estimation error is reduced iteratively. As we want to reduce the disturbance estimation error of agent$\#(j)$ (i.e. $e_{d,j}$) compared to the disturbance estimation error of agent$\#(j-1)$ (i.e. $e_{d,j-1}$), we will first establish a relationship between $e_{d,j}$ and $e_{d,j-1}$.

\subsection{Establishing relationship between $e_{d,j}$ and $e_{d,j-1}$:}\vspace{-7pt}
Let us first define a few system parameters. Based on the block diagram in Fig. \ref{fig:system_transfer_function2}, $G_{d,j}$ (dynamics from disturbance $d_j$ to output $z_{s,j}$), $G_{f,j}$ (dynamics from learning signal $d_{f,j}$ to output $z_{s,j}$), and $\Omega_j$ (dynamics from disturbance $d_j$ to DOB disturbance estimate $\hat{d}'_j$) can be described by: 
\begin{small}
 \begin{equation}
    \label{equation:Gdj}
     G_{d,j} = [1-Q_j+P(1)_j(M_j + C_j)]^{-1}P(1)_j(1-Q_j)
 \end{equation} 
 \begin{equation}
    \label{equation:Gfj}
    G_{f,j} = [1-Q_j+P(1)_j(M_j + C_j)]^{-1}(-P(1)_j)
\end{equation} 
\begin{equation}
    \label{equation:omegaj}
    \Omega_j=[1-Q_j+P(1)_j(M_j+C_j)]^{-1}\cdot (M_j+Q_j C_j) P(1)_j
\end{equation}
\end{small}

Now, using Fig. \ref{fig:system_transfer_function2} and Eq. (\ref{equation:omegaj}), the error in disturbance estimation of agent$\#(j)$ can be given by:
\begin{small}
\begin{equation}
    \label{equation:edj}
    \begin{split}
        e_{d,j}&=d_j-\hat{d}_j=d_j-(\hat{d}'_j+d_{f,j}) = (1-\Omega_j)\{d_j\}-d_{f,j}
    \end{split}
\end{equation}
\end{small}

Similarly, $e_{d,j-1}$ can be described by:
\begin{small}
\begin{equation}
    \label{equation:edj-1}
    e_{d,j-1}=(1-\Omega_{j-1})\{d_{j-1}\}-d_{f,j-1}
\end{equation}
\end{small}

Subtracting Eq. (\ref{equation:edj}) with Eq. (\ref{equation:edj-1}), we get:
\begin{small}
\begin{equation}
    \label{equation:relation1}
    \begin{split}
        e_{d,j}-e_{d,j-1}=(1-\Omega_j)\{d_j\}-&(1-\Omega_{j-1})\{d_{j-1}\}-d_{f,j}+d_{f,j-1}
    \end{split}
\end{equation}
\end{small}

Now, using Eq. (\ref{equation:zr_eq}), we replace $d_j$ with a term containing $d_{j-1}$ and rewrite Eq. (\ref{equation:relation1}) as:
\begin{small}
\begin{equation}
    \label{equation:relation2}
    \begin{split}
        e_{d,j}-e_{d,j-1}=&\left[(1-\Omega_j)\frac{\delta_j}{\delta_{j-1}}-(1-\Omega_{j-1})\right]\{d_{j-1}\}\\&-d_{f,j}+d_{f,j-1}
    \end{split}
\end{equation}
\end{small}

Now, Let us define a learning signal $d_{f,j}$ using the to-be-designed learning filters $L_{1,j}$ and $L_{2,j}$ as:
\begin{small}
\begin{equation}
    \label{equation:dfj}
    d_{f,j}=L_{1,j}\{e_{j-1}\}+L_{2,j}\{d_{f,j-1}\}
\end{equation}
\end{small}

Using Eq. (\ref{equation:dfj}) and with some simplifications, Eq. (\ref{equation:relation2}) can be written as:
\begin{small}
\begin{equation}
    \label{equation:relation3}
    \begin{split}
        e_{d,j}-e_{d,j-1}=\left[(1-\Omega_j)\frac{\delta_j}{\delta_{j-1}}-(1-\Omega_{j-1})\right]\{d_{j-1}\}\\-L_{1,j}\{e_{j-1}\}+(1-L_{2,j})\{d_{f,j-1}\}
    \end{split}
\end{equation}
\end{small}

Now, using Eq. (\ref{equation:Gdj}) and Eq. (\ref{equation:Gfj}), the displacement of the sprung mass for agent$\#(j-1)$ can be given by:
\begin{small}
\begin{equation}
    z_{s,j-1}=G_{d,j-1}\{d_{j-1}\}+G_{f,j-1}\{d_{f,j-1}\}
\end{equation}
\end{small}

Hence, with the aim of $z_{s,j}=0$, the error in trajectory tracking can be described as:
\begin{small}
\begin{equation}
    \label{equation:ej}
    e_{j-1}=-z_{s,j-1}=-G_{d,j-1}\{d_{j-1}\}-G_{f,j-1}\{d_{f,j-1}\}
\end{equation}
\end{small}
Using Eq. (\ref{equation:ej}), we can simplify Eq. (\ref{equation:relation3}) as:
\begin{small}
\begin{equation}
    \label{equation:relation4}
    \begin{split}
        e_{d,j}-&e_{d,j-1}=\left[(1-\Omega_j)\frac{\delta_j}{\delta_{j-1}}-(1-\Omega_{j-1})+L_{1,j}Gd_{j-1}\right]\{d_{j-1}\}\\&+(1-L_{2,j}+L_{1,j}Gf_{j-1})\{d_{f,j-1}\}
    \end{split}
\end{equation}
\end{small}

Now, using Eq. (\ref{equation:edj-1}), we substitute $d_{j-1}$ in Eq. (\ref{equation:relation4}) and simplify the Eq. (\ref{equation:relation4}) as:
\begin{small}
\begin{equation}
\label{equation:relation5}
    \begin{split}
        &e_{d,j}=\left[\frac{(1-\Omega_j)}{(1-\Omega_{j-1})}\frac{\delta_j}{\delta_{j-1}}+\frac{L_{1,j}Gd_{j-1}}{(1-\Omega_{j-1})}\right]\{e_{d,j-1}\}\\&+\left[\frac{(1-\Omega_j)}{(1-\Omega_{j-1})}\frac{\delta_j}{\delta_{j-1}}+\frac{L_{1,j}Gd_{j-1}}{(1-\Omega_{j-1})}-L_{2,j}+L_{1,j}Gf_{j-1}\right]\{d_{f,j-1}\}
    \end{split}
\end{equation}
\end{small}
For simplicity, let us define new terms $T_{e_{1,j}}$ and $T_{e_{2,j}}$ as:
\begin{small}
\begin{equation}
    \label{equation:te1j_te2j}
    \begin{split}
        T_{e_{1,j}}&=\frac{(1-\Omega_j)}{(1-\Omega_{j-1})}\frac{\delta_j}{\delta_{j-1}}+\frac{L_{1,j}Gd_{j-1}}{(1-\Omega_{j-1})}\\
        T_{e_{2,j}}&=\frac{(1-\Omega_j)}{(1-\Omega_{j-1})}\frac{\delta_j}{\delta_{j-1}}+\frac{L_{1,j}Gd_{j-1}}{(1-\Omega_{j-1})}-L_{2,j}+L_{1,j}Gf_{j-1}
    \end{split}
\end{equation}
\end{small}

Using Eq. (\ref{equation:te1j_te2j}), the Eq. (\ref{equation:relation5}) can be written as:
\begin{small}
\begin{equation}
    \label{equation:relation6}
    e_{d,j}= T_{e_{1,j}}\{e_{d,j-1}\}+T_{e_{2,j}}\{d_{f,j-1}\}
\end{equation}
\end{small}

Eq. (\ref{equation:relation6}) is the desired relationship between $e_{d,j}$ and $e_{d,j-1}$ for designing the learning filters. In subsection \ref{subsection:theorem}, we will introduce the proposed learning filters in a theorem and prove the error reduction using this relationship.
\subsection{Designing the learning filters}\vspace{-8pt}
We will define a few notations to be used in the theorem. Let us define an error reduction factor $0<\alpha<1$, where
\begin{equation}
    ||e_{d,j}||=\alpha \cdot ||e_{d,j-1}||
\end{equation}

and let us define $\eta_j$ as zero order approximation of a transfer function $\delta_j/\delta_{j-1}$. Also, for determining the learning filters for any agent, we will use the estimated plant models. These plant models may not be perfect due to modeling errors or variations in the model parameters. To quantify this uncertainty, we define $\Delta_{1,j}$ and $\Delta_{2,j}$ as: 
\begin{small}
\begin{equation}
    \label{equation:delta1j}
    P_j(1)=(1+\Delta_{1,j})\hat{P}_j(1)
\text{ and }
    P_j(2)=(1+\Delta_{2,j})\hat{P}_j(2)
\end{equation}
\end{small}
with $\hat{P}_j(1)$ and $\hat{P}_j(2)$ being the estimated plant models. $\Delta_{1,j}$ and $\Delta_{2,j}$ are small gain transfer functions to account for inaccuracies in these estimated plant models. From hereon, the hat symbol ($\hat{\cdot}$) over any transfer function indicates it is the estimated transfer function of the corresponding actual transfer function. With these notations, let us introduce the learning filters and prove the error reduction in the following theorem:

\label{subsection:theorem}
\textbf{Theorem:} With the learning filters:
\begin{small}
\begin{equation}
    \label{equation:L1j}
     L_{1,j}=\hat{G}_{d,j-1}^{-1}\left[\alpha(1-\hat{\Omega}_{j-1})-\eta_j({1-\hat\Omega}_j)\right] \text{ and}
\end{equation}
\begin{equation}
    \label{equation:L2j}
      L_{2,j}=\alpha+L_{1,j}\hat{G}_{f,j-1}
\end{equation}
\end{small}
we can achieve iterative estimation error reduction of factor $\approx \alpha$ in each iteration, where $0<\alpha<1$

\textbf{Proof:} Using Eq. (\ref{equation:L1j}) and Eq. (\ref{equation:L2j}), the Eq. (\ref{equation:te1j_te2j}) can be written as:
\begin{small}
\begin{equation}
    \label{equation:te1j_te2j_with_filters}
    \begin{split}
        T_{e_{1,j}}&=\frac{(1-\Omega_j)}{(1-\Omega_{j-1})}\frac{\delta_j}{\delta_{j-1}}+\frac{\hat{G}_{d,j-1}^{-1}\left[\alpha(1-\hat{\Omega}_{j-1})-\eta_j({1-\hat\Omega}_j)\right]Gd_{j-1}}{(1-\Omega_{j-1})}\\
        T_{e_{2,j}}&=\frac{(1-\Omega_j)}{(1-\Omega_{j-1})}\frac{\delta_j}{\delta_{j-1}}+\frac{\hat{G}_{d,j-1}^{-1}\left[\alpha(1-\hat{\Omega}_{j-1})-\eta_j({1-\hat\Omega}_j)\right]Gd_{j-1}}{(1-\Omega_{j-1})}\\&-(\alpha+L_{1,j}\hat{G}_{f,j-1})+L_{1,j}Gf_{j-1}
    \end{split}
\end{equation}
\end{small}
Now, Let us expand Eq. (\ref{equation:te1j_te2j_with_filters}) using Eq. (\ref{equation:Gdj}), Eq. (\ref{equation:Gfj}), Eq. (\ref{equation:omegaj}), and Eq. (\ref{equation:delta1j}). Also, let us assume that the low pass filter for all the systems is the same (i.e. $Q_j\approx Q$ for $\forall j$). With this, Eq. (\ref{equation:te1j_te2j_with_filters}) can be expressed as:
\begin{small}
\begin{equation}
\label{equation:Te1j_Te2j_intermediate}
    \begin{split}
        &T_{e_{1,j}}= \left(1+\frac{\Delta_{1,j-1}Q}{1+C_{j-1}P_{j-1}(1)}\right)\cdot\\&\left[\frac{(\alpha-\eta_j)\left(1+\frac{\Delta_{1,j-1}}{1+C_{j-1}P_{j-1}(1)}\right)}{\left(1+\frac{\Delta_{1,j-1}Q}{1+C_{j-1}P_{j-1}(1)}\right)}\right.\left. +\frac{\delta_j}{\delta_{j-1}} \frac{\frac{(1+\Delta_{1,j-1})(1+\Delta_{2,j})}{(1+\Delta_{1,j})(1+\Delta_{2,j-1})}}{\left(1+\frac{\Delta_{1,j}Q}{1+C_{j}P_{j}(1)}\right)}\right]\\&
        \\&
        T_{e_{2,j}}=T_{e_{1,j}}-\eta_j-(\alpha-\eta_j)\left(1+\frac{\Delta_{1,j-1}}{1+C_{j-1}P_{j-1}(1)}\right)
    \end{split}
\end{equation}
\end{small}
In Eq. (\ref{equation:Te1j_Te2j_intermediate}), any term of the form \begin{small}$1/(1+C_{j}P_{j}(1))$\end{small} reflects the transfer function from reference to the trajectory tracking error for agent$\#(j)$. For a well designed baseline active suspension controller, this term will have a gain very close to zero. Additionally, \begin{small}$\Delta_{1,j}/(1+C_{j}P_{j}(1))$\end{small} and \begin{small}$(\Delta_{1,j}Q)/(1+C_{j}P_{j}(1))$\end{small} will have even smaller gain, hence:
\begin{small}
    \begin{equation}
        \label{equation:closeto1_1}
        \left(1+\frac{\Delta_{1,j}}{1+C_{j}P_{j}(1)}\right)\approx1 \text{ and}
        \left(1+\frac{\Delta_{1,j}Q}{1+C_{j}P_{j}(1)}\right)\approx1 \text{ for } \forall j
    \end{equation}
\end{small}
Also, we can assume 
\begin{small}
\begin{equation}\label{equation:closeto1_3}\frac{(1+\Delta_{1,j-1})(1+\Delta_{2,j})}{(1+\Delta_{1,j})(1+\Delta_{2,j-1})}\approx 1
\end{equation} 
\end{small}

as $\Delta_{1,j}$ and $\Delta_{2,j}$ have a small gain for reasonably well modelled plants for $\forall j$

Using Eq. (\ref{equation:closeto1_1}), and Eq. (\ref{equation:closeto1_3}), the Eq. (\ref{equation:te1j_te2j_with_filters}) reduces to:
\begin{small}
\begin{equation}
    \label{equation:te1j_te2j_reduced}
    \begin{split}
    T_{e_{1,j}}\approx\alpha+\left(\frac{\delta_j}{\delta_{j-1}}-\eta_j\right)\frac{(1-\Omega_j)}{(1-\Omega_{j-1})}\\
    T_{e_{2,j}}\approx\left(\frac{\delta_j}{\delta_{j-1}}-\eta_j\right)\frac{(1-\Omega_j)}{(1-\Omega_{j-1})}
    \end{split}
\end{equation}
\end{small}
For suspension systems, $\delta_j/\delta_{j-1}$ can be approximated as a static gain for desired road profile frequency range. As $\eta_j$ is zero-order approximation of the transfer function $\delta_j/\delta_{j-1}$, $\delta_j/\delta_{j-1}-\eta_j\approx0$. Hence, Eq. (\ref{equation:te1j_te2j_reduced}) can be simplified as:
\begin{equation}
    \label{equation:te1j_te2j_reduced_further}
    \begin{split}
    T_{e_{1,j}}\approx\alpha \textbf{ and }
    T_{e_{2,j}}\approx0
    \end{split}
\end{equation}

Using Eq. (\ref{equation:te1j_te2j_reduced_further}), Eq. (\ref{equation:relation6}) can be written as:
\begin{equation}
    e_{d,j}\approx \alpha\{e_{d,j-1}\}
\end{equation}

Hence, the estimation error will be reduced by a factor $\approx \alpha$ in each iteration. This is the end of the theorem proof.

\begin{figure}[!t]
     \centering \includegraphics[width=0.43\textwidth]{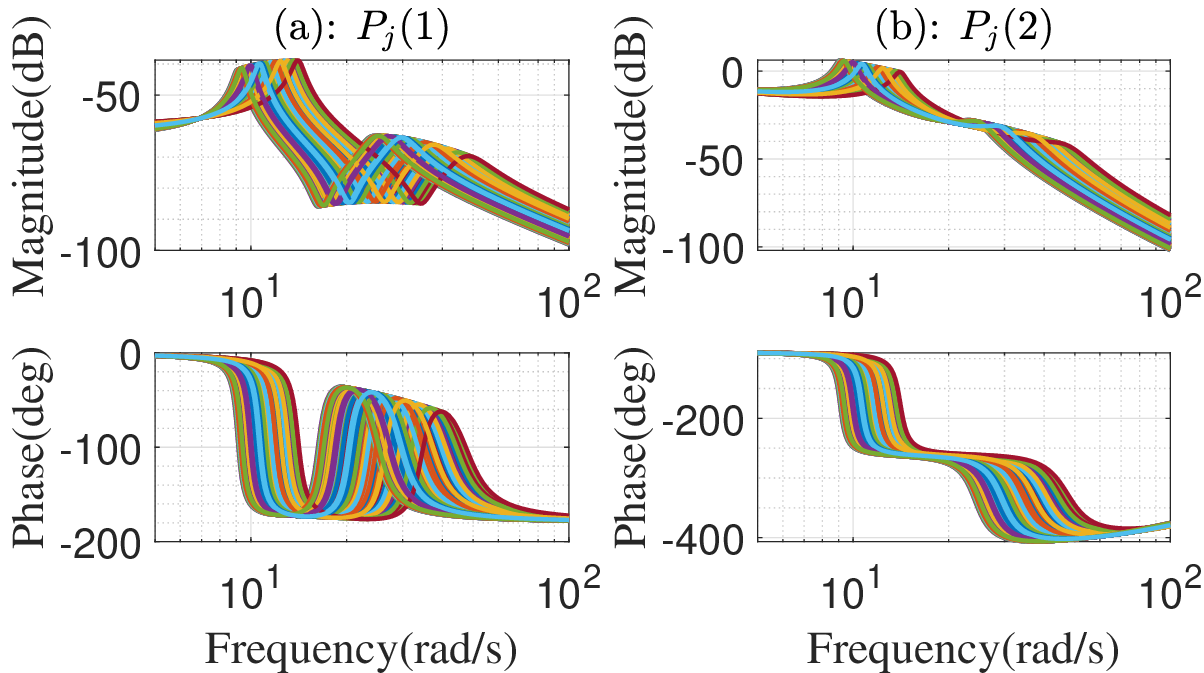}
  \caption{Bode plots of all the agents for transfer functions from (a) active suspension force to sprung mass displacement (i.e. $P_j(1)$) and (b) from road profile to sprung mass displacement (i.e. $P_j(2)$) }
  \label{fig:bode_plots}
\end{figure}

\section{Numerical Evaluation}
\label{section:simulation}
We evaluate the performance of the learning framework via numerical studies. We perform the simulation over 90 dynamically different agents running on the same road profile in succession. We use a quarter-car model of the vehicle suspension system as we had illustrated in Fig. \ref{fig:framework}. The specific characteristics of the agents used in the simulation are described in Table \ref{table:parameters}. The definitions used to generate both actual and nominal models are shown in the table. The simulator uses the actual model to simulate the response of the system, but the learning framework has access to an inaccurate nominal model only. This ensures the numerical evaluation reflects modeling uncertainties described by $\Delta_{1,j}$ and $\Delta_{2,j}$ in the theorem.

\begin{figure}[!htpb]
 \centering 
\includegraphics[width=0.41\textwidth]{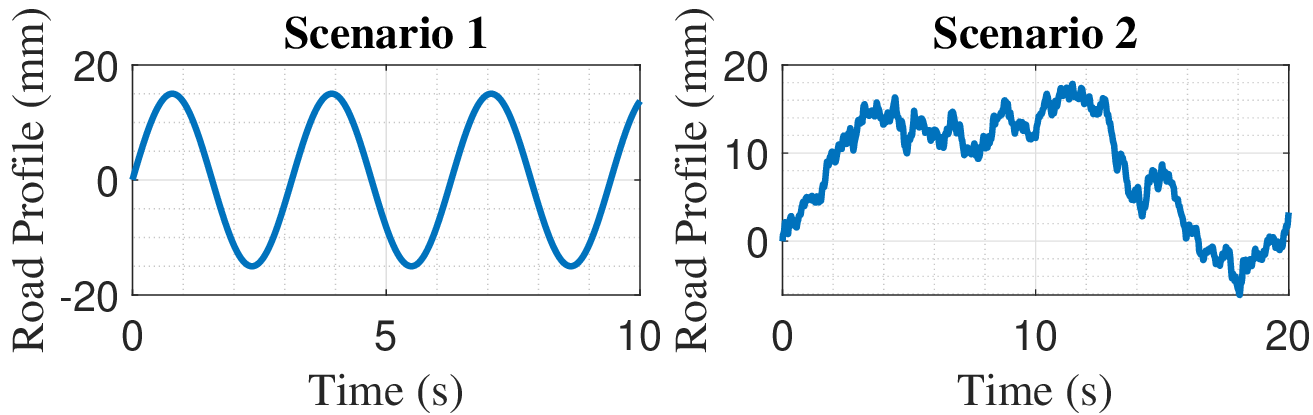}
\caption{Road profiles introduced}
\label{fig:road_profiles}
\end{figure}

\begin{figure*}[!htpb]
         \centering    \includegraphics[width=0.85\textwidth]{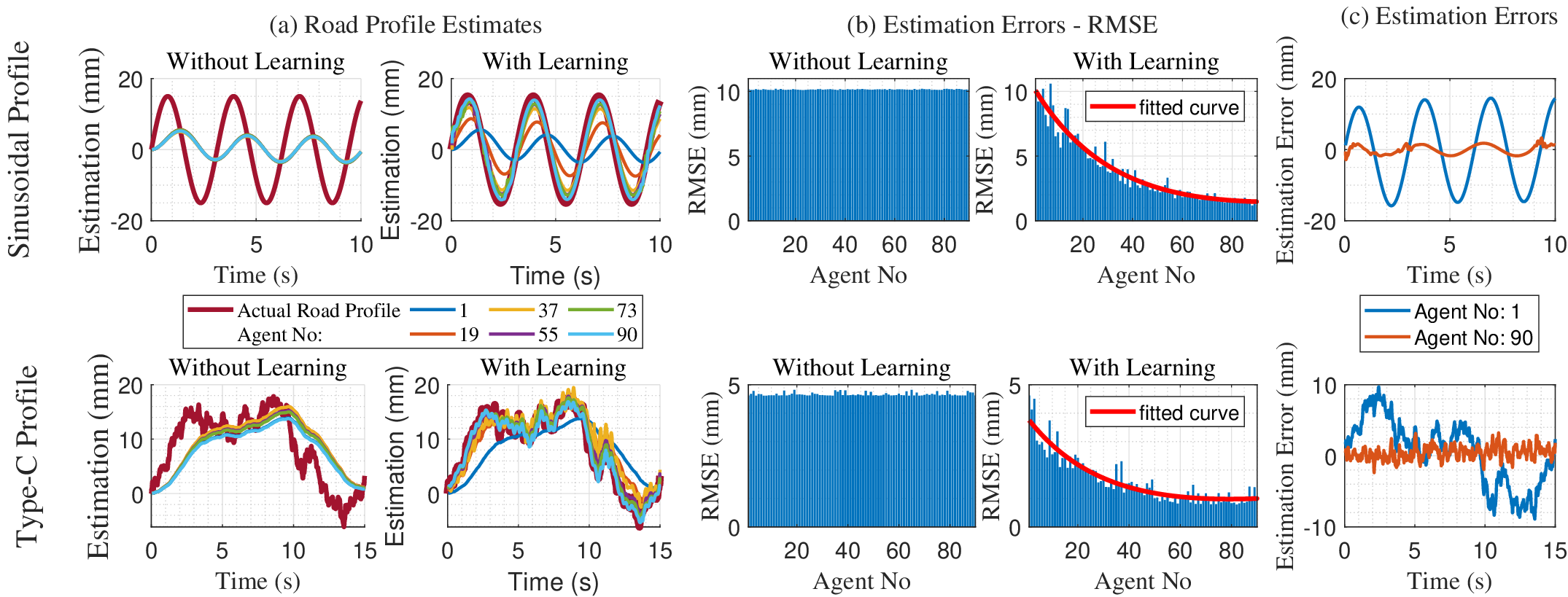}
      \caption{(a) Comparison of road profile estimates for a few agents for a sinusoidal road profile [top] and type-C road profile [bottom](b) Comparison of ``Root Mean Square Errors" of the road profile estimations for all agents for sinusoidal road profile [top] and type-C road profile [bottom] (c) Comparison of road profile estimation errors between the first and the last agents with learning for sinusoidal road profile [top] and type-C road profile [bottom]}
      \label{fig:numerical_results}
\end{figure*}

Each agent's dynamics is varied using the index multiplier $j$ varying from 1 to 90. In our current simulations, we choose $\beta=1/15$. Hence, for example, the actual sprung mass varies in the range of $2.45+1/15\times0=2.45\ kg$ to $2.45+1/15\times90=8.45\ kg$. We also randomize the order of $j$ to ensure that none of the characteristics steadily increase or decrease with each successive agent. Each $f_x$ (e.g. $f_{m_s}$, $f_{c_{us}}$, etc.) indicates the relevant model uncertainty added to defer the nominal model from the actual model. The $f_x$ is chosen as a random number with a mean of $1.0$ and an upper bound chosen before starting the simulation. Importantly, $f_x$ is not known to the learning framework. In our current setup, we have chosen this upper bound on uncertainty as $10\%$. Fig. \ref{fig:bode_plots} (a) and Fig. \ref{fig:bode_plots} (b) show the bode plots of actual model $P_j(1)$ and $P_j(2)$ respectively for all the agents. For each of our agents, we are using a PID controller as the baseline controller. Iterating again, the learning framework does not require any specific controller type as long as it is stable. Unlike the agent plant models, we do not need to add uncertainties to the controller as the controller is known to us.

\begin{table}[!htpb]
\centering
\begin{tabular}{|c|c|c|}
    \hline
    Parameter & Actual Model & Nominal Model \\
    \hline
    $m_s$ (kg)& $(2.45+\beta \times j)$ & $(2.45+\beta \times j)f_{m_{s}}$\\
    $m_{us}$ (kg) & $(1+\beta \times j)$ & $(1+\beta \times j)f_{m_{us}}$\\
    $k_s$ (N/m)& $(950+100\beta \times j)$ & $(950+100\beta \times j)f_{k_s}$\\
    $k_{us}$ (N/m)& $(1250+100\beta \times j)$ & $(1250+100\beta \times j)f_{k_{us}}$\\
    $c_s$ (N s/m)& $(7.5+\beta \times j)$ & $(7.5+\beta \times j)f_{c_s}$\\
    $c_{us}$ (N s/m)& $(5+\beta \times j)$ & $(5+\beta \times 7j)f_{c_{us}}$\\
    \hline
    Controller P Gain& \multicolumn{2}{c|}{$(1500+ 30\beta\times j)$}\\
    Controller I Gain& \multicolumn{2}{c|}{$(200+\beta \times j)$}\\
    Controller D Gain& \multicolumn{2}{c|}{$(500+ 15\beta \times j)$}\\
    \hline
\end{tabular}
\vspace{7pt}
\caption{Simulation agent characteristics. In current simulations, $\beta=1/15$}
\label{table:parameters}
\vspace{-15pt}
\end{table}

We perform simulation over 2 different road profile scenarios as shown in Fig. \ref{fig:road_profiles}. In scenario 1, a sinusoidal road profile of magnitude $15\ mm$ and frequency $5\ rad/s$ is used. In scenario 2, we introduce a type-C road profile [\cite{previous1}]. We also perform the simulations without learning but with DOB for all the agents to compare the results of the ``no learning" cases with the ``learning" cases. Fig. \ref{fig:numerical_results} shows the simulation results for both scenarios, with the top row for the sinusoidal road profile and the bottom row for the type-C road profile. 

\begin{figure}[!htpb]
 \centering    \includegraphics[width=0.4\textwidth]{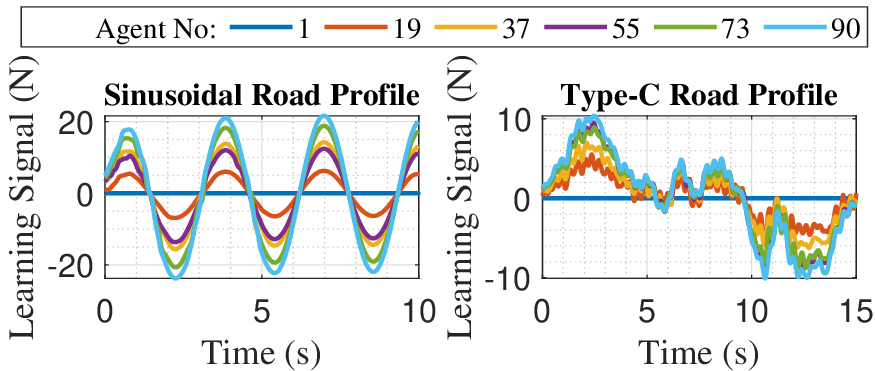} 
\caption{Learning signals for a few agents for (a) sinusoidal road profile and (b) Type-C Road Profile}
\label{fig:learning_signals}
\end{figure}

Fig. \ref{fig:numerical_results}(a) compares the road profile estimations between ``without learning" and ``with learning" cases. We show only a few agents' estimates out of 90 agents to de-clutter the plots. The actual road profile is also shown in all the plots. Without learning, the estimates from all the agents are almost similar, having some variations due to differences in the vehicle models. These estimates have a significant delay compared to the actual road profile as the DOB uses the delayed outputs of the system for estimation. These estimates are also scaled-down compared to the actual road profile in the sinusoidal profile case and missing high-frequency components in the type-C profile case. With learning, the road profile estimates iteratively approach the actual road profile, making agent$\#(90)$'s estimates very close to the actual road profile. 

Fig. \ref{fig:numerical_results}(b) shows the ``Root Mean Square Error" (RMSE) of road profile esitmates for all agents in both scenarios. This plot represents the overall ``closeness" of the road profile estimates to the actual road profiles. Without learning, all the agents have almost the same RMSE of around $10.25\ mm$ for the sinusoidal and around $4\ mm$ for the type-C road profiles. With learning, the errors decrease exponentially in each iteration. The RMSE converges to around $1.48\ mm$ and $0.98\ mm$ for sinusoidal and type-C road profiles respectively. 

Fig. \ref{fig:numerical_results} (c) shows the road profile estimation errors for agent$\#(1)$ (i.e. the first agent) and agent$\#(90)$ (i.e. the last agent) in the learning cases. It can be confirmed that the estimation errors for the last agent are negligible compared to the first agent in both scenarios.

Fig. \ref{fig:learning_signals} shows the learning signals generated by the learning framework for a few agents. These plots represent how the iterative learning process corrects inaccuracies of DOB. The learning signal for the agent$\#(1)$ is a constant $0$ as there is no previous agent to learn from. Agent$\#(2)$ onwards, the learning signal tries to compensate for the errors of DOB. The learning signals slowly converge toward the learning signal of the last agent (i.e. agent$\#(90)$).

\section{Conclusion}
\label{section:conclusion}
\vspace{-7pt}
This paper established a learning framework for estimating the road profile using DOB and iterative learning control using the onboard sensors. The framework utilizes multiple dynamically different active suspension vehicles to estimate the road profile accurately simultaneously increasing the passenger comfort in each vehicle. The learning framework can incorporate modeling uncertainties to iteratively reduce road profile estimation errors. The framework has been evaluated in robust simulations, confirming the learning process' effectiveness. 
\vspace{-4pt}
\bibliography{ifacconf}             

\begin{thebibliography}{24}
\providecommand{\natexlab}[1]{#1}
\providecommand{\url}[1]{\texttt{#1}}
\providecommand{\urlprefix}{URL }
\expandafter\ifx\csname urlstyle\endcsname\relax
  \providecommand{\doi}[1]{doi:\discretionary{}{}{}#1}\else
  \providecommand{\doi}{doi:\discretionary{}{}{}\begingroup \urlstyle{rm}\Url}\fi

\bibitem[{Chen et~al.(2022)Chen, Hajidavalloo, Li, and Zheng}]{chen2022cascaded}
Chen, Z., Hajidavalloo, M.R., Li, Z., and Zheng, M. (2022).
\newblock A cascaded learning framework for road profile estimation using multiple heterogeneous vehicles.
\newblock \emph{Journal of Dynamic Systems, Measurement, and Control}, 144(10), 104501.

\bibitem[{Chen et~al.(2020)Chen, Liang, and Zheng}]{chen2020knowledge}
Chen, Z., Liang, X., and Zheng, M. (2020).
\newblock Knowledge transfer between different uavs for trajectory tracking.
\newblock \emph{IEEE Robotics and Automation Letters}, 5(3), 4939--4946.

\bibitem[{Doumiati et~al.(2011)Doumiati, Victorino, Charara, and Lechner}]{contact_method2}
Doumiati, M., Victorino, A., Charara, A., and Lechner, D. (2011).
\newblock Estimation of road profile for vehicle dynamics motion: experimental validation.
\newblock In \emph{Proceedings of the 2011 American control conference}, 5237--5242. IEEE.

\bibitem[{Frej et~al.(2023)Frej, Moreau, Guridis, Benine-Neto, and Hernette}]{hinf_onboard}
Frej, G.B.H., Moreau, X., Guridis, R., Benine-Neto, A., and Hernette, V. (2023).
\newblock Road profile estimation from onboard sensor measurements through a combination of h-infinity and unknown inputs observers.
\newblock In \emph{2023 31st Mediterranean Conference on Control and Automation (MED)}, 113--118.
\newblock \doi{10.1109/MED59994.2023.10185895}.

\bibitem[{Gao et~al.(2020)Gao, Li, and Wang}]{gao2020privacy}
Gao, H., Li, Z., and Wang, Y. (2020).
\newblock Privacy-preserved collaborative estimation for networked vehicles with application to road anomaly detection.
\newblock \emph{arXiv preprint arXiv:2008.02928}.

\bibitem[{G{\"o}hrle et~al.(2014)G{\"o}hrle, Schindler, Wagner, and Sawodny}]{active_sensor_front}
G{\"o}hrle, C., Schindler, A., Wagner, A., and Sawodny, O. (2014).
\newblock Road profile estimation and preview control for low-bandwidth active suspension systems.
\newblock \emph{IEEE/ASME Transactions on Mechatronics}, 20(5), 2299--2310.

\bibitem[{Hajidavalloo et~al.(2022)Hajidavalloo, Li, Xia, Louati, Zheng, and Zhuang}]{Mohammad_TITS}
Hajidavalloo, M.R., Li, Z., Xia, X., Louati, A., Zheng, M., and Zhuang, W. (2022).
\newblock Cloud-assisted collaborative road information discovery with gaussian process: Application to road profile estimation.
\newblock \emph{IEEE Transactions on Intelligent Transportation Systems}, 23(12), 23951--23962.
\newblock \doi{10.1109/TITS.2022.3194093}.

\bibitem[{Hassen et~al.(2019)Hassen, Miladi, Abbes, Baslamisli, Chaari, and Haddar}]{previous1}
Hassen, D.B., Miladi, M., Abbes, M.S., Baslamisli, S.C., Chaari, F., and Haddar, M. (2019).
\newblock Road profile estimation using the dynamic responses of the full vehicle model.
\newblock \emph{Applied Acoustics}, 147, 87--99.

\bibitem[{Healey et~al.(1977)Healey, Nathman, and Smith}]{contact_method1}
Healey, A.J., Nathman, E., and Smith, C.C. (1977).
\newblock {An Analytical and Experimental Study of Automobile Dynamics With Random Roadway Inputs}.
\newblock \emph{Journal of Dynamic Systems, Measurement, and Control}, 99(4), 284--292.
\newblock \doi{10.1115/1.3427121}.
\newblock \urlprefix\url{https://doi.org/10.1115/1.3427121}.

\bibitem[{Li et~al.(2017)Li, Kolmanovsky, Atkins, Lu, Filev, and Bai}]{comfort}
Li, Z., Kolmanovsky, I.V., Atkins, E.M., Lu, J., Filev, D.P., and Bai, Y. (2017).
\newblock Road disturbance estimation and cloud-aided comfort-based route planning.
\newblock \emph{IEEE Transactions on Cybernetics}, 47(11), 3879--3891.
\newblock \doi{10.1109/TCYB.2016.2587673}.

\bibitem[{Li et~al.(2019)Li, Zheng, and Zhang}]{previous2}
Li, Z., Zheng, M., and Zhang, H. (2019).
\newblock Optimization-based unknown input observer for road profile estimation with experimental validation on a suspension station.
\newblock In \emph{2019 American Control Conference (ACC)}, 3829--3834. IEEE.

\bibitem[{Ma et~al.(2020)Ma, Wang, Yang, and Yang}]{autonomous_vehicles_survey}
Ma, Y., Wang, Z., Yang, H., and Yang, L. (2020).
\newblock Artificial intelligence applications in the development of autonomous vehicles: a survey.
\newblock \emph{IEEE/CAA Journal of Automatica Sinica}, 7(2), 315--329.
\newblock \doi{10.1109/JAS.2020.1003021}.

\bibitem[{Massaro et~al.(2016)Massaro, Ahn, Ratti, Santi, Stahlmann, Lamprecht, Roehder, and Huber}]{cars_have_sensors}
Massaro, E., Ahn, C., Ratti, C., Santi, P., Stahlmann, R., Lamprecht, A., Roehder, M., and Huber, M. (2016).
\newblock The car as an ambient sensing platform [point of view].
\newblock \emph{Proceedings of the IEEE}, 105(1), 3--7.

\bibitem[{McCann and Nguyen(2007)}]{laser_method1}
McCann, R. and Nguyen, S. (2007).
\newblock System identification for a model-based observer of a road roughness profiler.
\newblock In \emph{2007 IEEE Region 5 Technical Conference}, 336--343. IEEE.

\bibitem[{Modi et~al.(2024)Modi, Chen, Liang, and Zheng}]{previous4}
Modi, H., Chen, Z., Liang, X., and Zheng, M. (2024).
\newblock Improving disturbance estimation and suppression via learning among systems with mismatched dynamics.
\newblock \emph{IEEE Robotics and Automation Letters}, 9(6), 5238--5245.
\newblock \doi{10.1109/LRA.2024.3391026}.

\bibitem[{Ni et~al.(2020)Ni, Li, Zhao, and Kong}]{laser_method2}
Ni, T., Li, W., Zhao, D., and Kong, Z. (2020).
\newblock Road profile estimation using a 3d sensor and intelligent vehicle.
\newblock \emph{Sensors}, 20(13), 3676.

\bibitem[{Peraka and Biligiri(2020)}]{pavement_assessment_for_decision_making}
Peraka, N.S.P. and Biligiri, K.P. (2020).
\newblock Pavement asset management systems and technologies: A review.
\newblock \emph{Automation in Construction}, 119, 103336.

\bibitem[{Rath et~al.(2014)Rath, Veluvolu, and Defoort}]{previous3}
Rath, J.J., Veluvolu, K.C., and Defoort, M. (2014).
\newblock Estimation of road profile for suspension systems using adaptive super-twisting observer.
\newblock In \emph{2014 European Control Conference (ECC)}, 1675--1680. IEEE.

\bibitem[{Sisi et~al.(2024)Sisi, Mirzaei, and Rafatnia}]{zus_displacement_estimation}
Sisi, Z.A., Mirzaei, M., and Rafatnia, S. (2024).
\newblock Estimation of vehicle suspension dynamics with data fusion for correcting measurement errors.
\newblock \emph{Measurement}, 114438.

\bibitem[{Song and Wang(2020)}]{active_MPC}
Song, S. and Wang, J. (2020).
\newblock Incremental model predictive control of active suspensions with estimated road preview information from a lead vehicle.
\newblock \emph{Journal of Dynamic Systems, Measurement, and Control}, 142(12), 121004.

\bibitem[{Tudón-Martínez et~al.(2015)Tudón-Martínez, Fergani, Sename, Martinez, Morales-Menendez, and Dugard}]{semi_active_Q_parameterization}
Tudón-Martínez, J.C., Fergani, S., Sename, O., Martinez, J.J., Morales-Menendez, R., and Dugard, L. (2015).
\newblock Adaptive road profile estimation in semiactive car suspensions.
\newblock \emph{IEEE Transactions on Control Systems Technology}, 23(6), 2293--2305.
\newblock \doi{10.1109/TCST.2015.2413937}.

\bibitem[{Whaiduzzaman et~al.(2014)Whaiduzzaman, Sookhak, Gani, and Buyya}]{vehicular_cloud_computing}
Whaiduzzaman, M., Sookhak, M., Gani, A., and Buyya, R. (2014).
\newblock A survey on vehicular cloud computing.
\newblock \emph{Journal of Network and Computer applications}, 40, 325--344.

\bibitem[{Yu et~al.(2023)Yu, Evangelou, and Dini}]{active_suspension_advances}
Yu, M., Evangelou, S., and Dini, D. (2023).
\newblock Advances in active suspension systems for road vehicles.
\newblock \emph{Engineering}.

\bibitem[{Zheng et~al.(2020)Zheng, Lyu, Liang, and Zhang}]{zheng2020generalized}
Zheng, M., Lyu, X., Liang, X., and Zhang, F. (2020).
\newblock A generalized design method for learning-based disturbance observer.
\newblock \emph{IEEE/ASME Transactions on Mechatronics}, 26(1), 45--54.

\end{thebibliography}
\end{document}